\documentclass[12pt]{article}
\begin{document}
\title{The Hamilton-Jacobi Theory, Quantum Mechanics and General Relativity}
\author{B.G. Sidharth\\
International Institute for Applicable Mathematics \& Information Sciences\\
Hyderabad (India) \& Udine (Italy)\\
B.M. Birla Science Centre, Adarsh Nagar, Hyderabad - 500 063 (India)}
\date{}
\maketitle
\begin{abstract}
The Hamilton-Jacobi theory of Classical Mechanics can be extended in a novel manner to systems which are fuzzy in the sense that they can be represented by wave functions. A constructive interference of the phases of the wave functions then gives us back Classical systems. In a suitable description this includes both Quantum Theory and General Relativity in the well known superspace formulation. However, there are several nuances which provide insight into these latter systems. All this is considered in this paper together with suitable generalization, to cascades of super universes.
\end{abstract}
\section{Introduction}
The Hamilton-Jacobi theory of Mechanics is well known \cite{gold}. Let us now look at this theory from a slightly different perspective which enables its generalization to situations which cannot be described purely classically. As is usual we start with the action integral 
\begin{equation}
I = \int L \left(\frac{dx}{dt}, x, t\right) dt\label{e1}
\end{equation}
and extremalize it \cite{wheeler,mwt}. Let us denote, working for the moment in one space dimension for simplicity, the extremum integral of (\ref{e1}) by
\begin{equation}
S (x,t) = \bar{I}\label{e2}
\end{equation}
We then have,
\begin{equation}
p = \frac{\partial S}{\partial x}\label{e3}
\end{equation}
\begin{equation}
E = - \frac{\partial S}{\partial t} = \frac{p^2}{2m} + V (x)\label{e4}
\end{equation}
where $p$ denotes the momentum and $E$ denotes the energy. We can combine (\ref{e3}) and (\ref{e4}) and write,
\begin{equation}
E = H (p,x)\label{e5}
\end{equation}
Equation (\ref{e5}) is well known in Classical Mechanics but let us now introduce a description in terms of the wave function
\begin{equation}
\psi = Re^{(\imath/\hbar )S(x,t)}\label{e6}
\end{equation}
where $R$ is slowly varying and $S$ is given by (\ref{e2}). We now get from (\ref{e6}) the Hamilton-Jacobi equation in one dimension
\begin{equation}
- \frac{\partial S}{\partial t} = H \left(\frac{\partial S}{\partial x} , x\right) = \frac{1}{2m} \left(\frac{\partial S}{\partial x}\right)^2 + V (x)\label{e7}
\end{equation}
whose solution is
\begin{equation}
S = (x,t) = - Et + S \left\{ 2m\left[E - Vjk\right]\right\}^{1/2} dx + S\label{e8}
\end{equation}
In view of the description given in (\ref{e6}), (\ref{e7}) and (\ref{e8}) take on a more general character in terms of the so called dynamical phase $S$ . Further as is well known, the requirement 
$$\frac{\partial S}{\partial E} = 0$$
which denotes a constructive interference in the phase of systems described by wave functions leads us back to a classical particle description as in (\ref{e4}). We will now apply the above theory to Quantum Mechanical systems and General Relativity.
\section{Quantum Theory}
In Quantum Theory, the description of the wave function as in (\ref{e6}) when substituted in the Schrodinger equation 
\begin{equation}
\imath \hbar \frac{\partial \psi}{\partial t} = - \frac{\hbar^2}{2m} \nabla^2 \psi + V \psi\label{e9}
\end{equation}
leads to
\begin{equation}
\frac{\partial \rho}{\partial t} + \vec{\nabla} \cdot (\rho \vec{v}) = 0\label{e10}
\end{equation} 
\begin{equation}
\frac{1}{\hbar} \frac{\partial S}{\partial t} + \frac{1}{2m} (\vec{\nabla}S)^2 + \frac{V}{\hbar^2} - \frac{1}{2m} \frac{\nabla^2 R}{R} = 0\label{e11}
\end{equation}
where $\rho = R^2,\vec{v} = \frac{\hbar}{m} \vec{\nabla} S \, \mbox{and}\, Q \equiv - \frac{\hbar^2}{2m} (\nabla^2 R/R)$.\\ Equation (\ref{e10}) is easily recognized as the equation of continuity and (\ref{e11}) as the Hamilton-Jacobi equation in three dimensions.  At this stage there is still no surprise because the Bohmian description given in (\ref{e10}) and (\ref{e11}) as arising from the Schrodinger equation (\ref{e9}) describes a Quantum Theory without observers, essentially a deterministic Quantum Theory \cite{bohm}.\\
However the surprising result comes when we start with the stochastic approach of Nelson, that is 
\begin{equation}
\Delta x^2 \approx a \Delta t\label{e12}
\end{equation}
Equation (\ref{e12}) describes a Brownian or Diffusion process and can be shown to lead back to the Hamilton-Jacobi equation (\ref{e11}) once it is realized that the right time derivative is not equal to the left time derivative \cite{cu,nottale}. Equally surprising is the fact that the Feynman path integral method \cite{fh,it,gip} requires an integrability condition, which is the same as (\ref{e12}) and this gives us back the Hamilton-Jacobi description (\ref{e7}) and (\ref{e8}). We would like to emphasize in both of these approaches, that we are dealing with an ensemble of paths and it is only a constructive interference which singles out a particle like path, whereas in the Bohm description or in Classical Physics, there is an actual trajectory.
\section{General Relativity}
We now come to Wheeler's description of the wave function of the universe in terms of superspace (Cf.ref. \cite{wheeler,mwt}). Superspace is a manofold such that a single point therein represents the whole of the geometry of three dimensional space. As is usual we denote such a three dimensional geometry by (\ref{e3}) . Treating this as a point we introduce the wave function like description (\ref{e6}), except that we are in superspace which is essentially a four dimensional manifold. We summarize this situation following Wheeler.
\begin{table}
\caption{Superspace compared with Particle Dynamics}
\begin{tabular}{|c|c|c|} \hline
Quality &  Particle & Superspace \\ \hline
Dynamical entity & Particle & Space\\
Descriptors of & $x,t$(``event'') & $^{(3)}G$(``3-geometry'')\\
momentary configuration & & \\
History & $x = x(t)$ & $^{(4)}G$(``4-geometry'')\\
History is a & Yes. Every point & Yes. Every spacelike\\
stock pile of & on world line gives & slide through\\
configurations? & a momentary configuration & $^{(4)}G$ gives\\
& or particle & a momentary configuration\\
& & of space\\
Dynamic arena & Spacetime & Superspace\\ 
& (totality of all & (totality of all\\
& points $x,t)$ & $^{(3)}GS$)\\ \hline
\end{tabular}
\end{table}
Interestingly it ha
s been shown by Stern that a long as we are dealing with Euclidean type three dimensional spaces with positive definite metric, superspace constitutes a manifold in the sense that each point therein has a neighborhood homeomorphic to an open set in a Banach space and two distinct points are separated by disjoint neighborhoods. This enables us to carry out the usual operations in superspace and we are led back using the principle of constructive interference to this time the ten field equations of  Einstein.\\
Thus the Hamilton-Jacobi theory leads both to Quantum Mechanics and to General Relativity in terms of the wave function description (\ref{e6}) together with constructive interference. However it must be borne in mind that in Quantum Mechanics we are dealing with the usual three dimensional space whereas for obtaining the Einstein's equations of General Relativity, we are using the four dimensional superspace.\\
We can generalize the concept of superspace by considering rather than a four dimensional manifold a set of three dimensional universes, without imposing any conditions of smoothness or even continuity. Let us consider as above, the three dimensional space as a "point". It has been argued in detail by the author \cite{cu} that at the scale of the radius of the universe, the behavior mimics the Quantum behavior but with a scaled up Planck constant  given by 
\begin{equation}
h_1 \sim 10^{93}\label{e13}
\end{equation}
(Cf. also \cite{gip}). It can then be argued that the radius of the universe corresponds to the Compton wavelength of a particle and is given by
\begin{equation}
R = \frac{h_1}{Mc}\label{e14}
\end{equation}
where $M$ is the mass of the universe. Indeed one can argue that these scaled Quantum effects arise due to the well known equation of gravitational orbits
\begin{equation}
\frac{GM}{L} \sim v^2\label{e15}
\end{equation}
where $v$ in (\ref{e15}) is the dispersion of the velocities.\\
We can now consider the universe to be a Gaussian wave packet \cite{uof} and get,
\begin{equation}
R \approx \frac{\sigma}{\sqrt{2}} \left(1 + \frac{h_1^2 T^2}{\sigma^4 M^2}\right)^{1/2} \left(\approx \frac{1}{\sqrt{2}} \frac{h_1 T}{\sigma M}\right)\label{e16}
\end{equation}
where $T$ denotes the age of the universe and $\sigma$ the dispersion in length $\sim R$. Remembering that $R \approx cT$, (\ref{e16}) actually reduces to equation (\ref{e14})! The width of the wave packet is the ``Compton'' length. Differentiation of (\ref{e16}) gives,
\begin{equation}
\dot{R} \approx \frac{h_1}{\sigma M} \approx \frac{h_1}{M\sigma^2} \cdot R \equiv HR\label{e17}
\end{equation}
Equation (\ref{e17}) resembles Hubble's law. We can show that this is indeed so, by using the well known relation,
$$R = \sqrt{N}l$$
where $N \sim 10^{80}$ is the number of elementary particles in the universe and $I$ is a typical Compton wavelength
\begin{equation}
H \approx \frac{c}{\sqrt{N}l} \approx \frac{Gm^3 c}{h^2}\label{e18}
\end{equation}
Not only does (\ref{e18}) give the correct value of the Hubble constant, but it is also Weinberg's ``mysterious'' empirical relation, giving the pion mass in terms of the Hubble constant or vice versa.\\
Thus the expansion of the universe is due to the decay of the wave packet.\\
Interestingly, from (\ref{e16}), using again $R \approx cT$, we can deduce that
\begin{equation}
Mc^2 \cdot T \approx h_1\label{e19}
\end{equation}
Equation (\ref{e19}) and (\ref{e14}) are the analogues of Heisenberg's Uncertainty Principle.\\
In fact we can push the above considerations even further. We could consider Quantum effects to the super universe which is now not Wheeler's superspace but a collection of universes. In this case as in the explanation for voids in the universe \cite{bgs}, we could consider the different universes to be different quantized levels in a super atom. This time we get for a typical distance $R$ of a level
\begin{equation}
\frac{GM^3 R}{h^2_1} \sim 10\label{e20}
\end{equation}
Using the value of the scaled Planck constant $R$ can be easily calculated to be the radius of the universe.\\
Finally, it may be pointed out that the "Gravitational" force between the universes can be seen from (\ref{e20}) to be $\sim \frac{\mbox{constant}}{r}$  where $r$ is now the "distance" between the universes.
\section{Remarks}
1. We note that there is a rationale for the four dimensionality considered above. It has already been showed by the author (Cf.ref.\cite{uof} for a detailed discussion) that at the resulution of the Planck scale the universe can be considered to be a one dimensional array of Planck oscillators \cite{psu}. A simple way to see this is to observe that if there are $N'$ such coherent oscillators, at the Planck scale, then the total extent $R$ is given by,
\begin{equation}
R = \sqrt{N'} l_P\label{e21}
\end{equation}
It has also been shown \cite{ijmpe} that $N' \sim 10^{120}$. So (\ref{e21}) gives us back the radius of the universe. So also the mass of the universe is given by
\begin{equation}
M = m_P \sqrt{N'}\label{e22}
\end{equation}
where $m_P \sim 10^{-5}gms$ is the Planck mass. Equation (\ref{e22}) gives us the correct mass of the universe.\\
However the scale of the elementary particles is the Compton scale and it has been argued that at this scale we encounter two dimensionality (Cf.ref.\cite{cu}). (In fact if there is a collection of ultra relativistic particles, then it is known that the centres of mass form a two dimensional disc \cite{moller}). Well outside the Compton scale we encounter three dimensionality as it manifests by inverse square Coulomb or gravitational forces. Furthermore we have
$$M = \rho R^3 , \rho \sim 10^{-29}gm/cc$$
$\rho$ being the average density of the universe, as we should have for three dimensions. If now the universe as a whole is treated as having the scaled up Compton scale extension then well beyond this scale we should have by the same reasoning an extra that is fourth space dimension. This in fact is Wheeler's superspace. Our variation has been that this four dimensional space has not been taken as a manifold, but rather as a set where neither smooth nor continuum properties are assumed.\\
In the above considerations it is interesting to note that the universe itself shows up as a black hole (Cf.ref.\cite{cu}). This is brought out by the fact that the radius of the universe resembles the Schwarzchild radius, that is
$$R \sim \frac{2GM}{c^2}$$
Furthermore the time taken by light to reach an observer at the centre from the periphery is also the time taken by light to traverse this distance in an equivalent black hole: This is the age of the universe.\\
2. In the above we have generalized the Feynman path integral and stochastic approaches, valid for elementary particles to the case of the entire universe treated as a wave packet as we saw in equations like (\ref{e14}), (\ref{e16}) etc. On the other hand these two approaches relying on (\ref{e12}) show that we cannot go down to arbitrarily small space time scales for then we would encounter infinite velocities. In fact as Feynman put it \cite{fh},\\
``... these irregularities are such that the ``average'' square velocity does not exist, where we have used the classical analogue in referring to an ``average.''\\
``If some average velocity is defined for a short time interval $\Delta t$, as, for example, $[x(t + \Delta t) - x(t)]/\Delta t$, the ``mean'' square value of this is $-\hbar /(\imath m \Delta t)$. That is, the ``mean'' square value of a velocity averaged over a short time interval is finite, but its value becomes larger as the interval becomes shorter.\\
It appears that quantum-mechanical paths are very irregular. However, these irregularities average out over a reasonable length of time to produce a reasonable drift, or ``average'' velocity, although for short intervals of time the ``average'' value of the velocity is very high...''\\
This immediately provides a rationale for considering a non smooth set (instead of Wheeler's superspace) as discussed in detail in reference \cite{gip}.\\
However it must be reiterated that in Wheeler's superspace approach General Relativity appears as a four space dimensional generalization of the Quantun wave function (\ref{e6}). This is again brought out in our approach where we have used the Gaussian wave packet for the universe as in (\ref{e16}) by the fact that (\ref{e20}) in fact leads to the Weinberg formula or equivalently to (\ref{e18}) giving the gravitational constant $G$ in terms of microphysical parameters $l,m,h$ and $N$ the number of particles in the whole of the universe. In other words gravitation appears due to not microphysical considerations but rather due to considerations at the scale of all the particles in the universe \cite{fpl}.\\
3. In the Gaussian packet for the universe, the decay which appears as a Hubble expansion as in (\ref{e17}) can also be thought as due to a time varying $G$ which follows from (\ref{e18}) as \cite{uzan}
\begin{equation}
|\dot{G}/G| \sim \frac{1}{T}\label{e23}
\end{equation}
The loss of energy due to (\ref{e23}) leads to the decay of the wave packet for in this case of the Gaussian packet \cite{uof} we have, the spread of all the $N$ elementary particles that is the spread of the universe given by
\begin{equation}
R = \frac{GNm}{c^2} \approx l \cdot \frac{\hbar t}{l^2m}\label{e24}
\end{equation}
In fact (\ref{e24}) is an alternative form of (\ref{e18}).\\
4. In fact a similar consideration as above can explain the decay of a Planck mass particle to an elementary particle in the Compton time (Cf.ref.\cite{gip}). The same arguments can now apply with scaled Schrodinger equationsto multiple universes.\\
5. The process by which the Planck oscillators condense into elementary particles which then constitute our universe can continue using the scaled Planck constants and we have a multiplicity of universes $u_\imath$ forming a super universe $\bar{u}$, say. There could be a multiplicity of super universes $\bar{u}j$ and so on - nothing that we know really forbids such a self similar indefinite construction. At each stage however we would be going to higher and higher dimensions, just as a super universe would require four space dimensions.\\
6. We finally make a few comments: For the past few years several astronomers and physicists have come round to the view that our universe is only a tiny constituent of a multiverse. In the words of David Deutsch of Oxford University \cite{david} ``A growing number of physicists, myself included, are convinced that the thing we call 'the universe'- namely space, with all the matter and energy it contains - is not the whole of reality. According to quantum theory - the deepest theory known to physics - our universe is only a tiny facet of a larger multiverse, a highly structured continuum containing many universes.''\\
This view replaces the earlier Wheeler concept of multiple universes connected by gravitational wormholes (Cf. discussion in ref.\cite{cu}). The more recent view arises from the big bang singularity and stems from the realization that at this point of singularity, it is impossible to predict physical behavior or physical laws. On the other hand our universe itself appears to be highly fine tuned in the sense that even a minute departure from the values of the fundamental constants would lead to a very different type of a universe - in particular one which may not be able to support any life at all. In this view, which is in the spirit of the anthropic principle, there would be several universes, and ours is merely one with the right parameters. As the well known astronomer Martin Rees put it, ``Our universe may be just one element, one atom as it were - in an infinite ensemble.... each universe starts with its own big bang, acquires a distinctive imprint (and its individual physical laws) as it cools, and traces out its own cosmic cycle. The big bang that triggered our entire universe is, in this grander perspective, an infinitesimal part of an elaborate structure that extends far beyond the range of any telescopes.''\\
In any case it has been realized that by thinking that our observable universe is all there is, we may be as naive as the pre-Copernican scholars who believed that the earth was at the centre of the universe. In fact it is being speculated that there may be something like $10^{500}$ universes like our own!\\
In our approach, we indicate how the concept of super universes can be described on a mathematical footing. Let us start with the Hamilton-Jacobi equation (\ref{e11}), but in four dimensions, a la Wheeler. At this point we depart and consider the generalized solinoidal velocity field 
\begin{equation}
\vec{\nabla} \cdot \vec{v} = 0\label{e25}
\end{equation}
As has been argued, we can still obtain a non zero circulation from (\ref{e25}), viz.,
\begin{equation}
\Gamma = \int_c \vec{v} \cdot \vec{d} r = (h_1 /m) \int_c \vec{\nabla} S \cdot d \vec{r} = (h_1 /m) \oint dS = \frac{\pi h_1 n}{m}, n = 1,2,\cdots\label{e26}
\end{equation}
In (\ref{e26}) $h_1$ is the scaled up Planck constant encountered before. In the usual theory, it has been argued that the analogue of (\ref{e26}) describes a multiply connected space and in fact is the explanation for Quantum Mechanical spin \cite{bgs2}. In the case of the universe at large (\ref{e26}) would be the counterpart, precisely the equation behind the scaled up Compton wavelength (\ref{e14}). This would also be the spin of the universe demonstrated by Godel for the Einstein equations. A cosmic footprint for these considerations may have already been found in the anisotropies of the Cosmic Microwave Background Radiation \cite{cu}. From this point of view each of the universes contained in the four dimensional super universe would be no more than the counterpart of the spinning elementary particle, not to mention the super super universes.\\
To press these considerations further we note that we can use the Beckenstein temperature formula for a black hole \cite{ruffini}
\begin{equation}
T = \frac{h c^3}{8\pi Gkm}\label{e27}
\end{equation}
If in (\ref{e27}), we replace $h$ by $h_1$ and the mass $m$ by $M$, the mass of the universe, then we get
$$kT = Mc^2$$
This is consistent, as the right side is the energy content of the universe.\\
Further the decay time in this black hole theory is given by
\begin{equation}
t = \frac{1}{3\beta} m^3 , \, \, \beta = \frac{h c^4}{(30.8)^3 \pi G^2}\label{e28}
\end{equation}
In (\ref{e28}) with the scaled up Planck constant and the mass of the universe we get, the age to be
\begin{equation}
t \sim 10^{21}sec\label{e29}
\end{equation}
Equation (\ref{e29}) shows that the universe would decay in a time span that is $\sim 10^4$ times its present age. On the other hand we could use in (\ref{e29}) the variation of the gravitational constant \cite{bgsfpl} viz.,
\begin{equation}
G = \frac{l c^2}{\sqrt{N}m} = \frac{l^2 c}{m t}\label{e30}
\end{equation}
Then with this factored in we find that (\ref{e29}) holds approximately.\\
We can then speak of the force between the different universes within the super universe. This would follow from a generalization of (\ref{e30}) which gives the gravitational force between the $N$ particles of the universe. If we take the number of universes in the super universe to be $\sim 10^{100}$, say, then the replacement in (\ref{e30}) of the cosmic parameters gives the analogue of the gravitational constant as
$$G' \sim 10^{-55}$$
This constant is some $10^{-47}$ times as weak as the gravitational constant itself.  

\end{document}